# Design of the ELIMAIA ion collection system

F Schillaci[a], G A P Cirrone[a,b], G Cuttone[a], M Maggiore[c], L Andò[a], A Amato[a], M Costa[a], G Gallo[a], G Korn[b], G La Rosa[a], R leanza[a], R Manna[a], D Margarone[b], G Milluzzo[a], S Pulvirenti[a], F Romano[a], S Salamone[a], M Sedita[a], V Scuderi[a,b], A Tramontana[a,d]

[a] *Istituto Nazionale di Fisica Nucleare, Laboratori Nazionali del Sud,*
 *Via S. Sofia 62, Catania, Italy*

[b] *Institute of Physics ASCR, v.v.i (FZU), ELI-Beamlines Project,*
 *Na Slovance 2, Prague, Czech Republic*

[c] *Istituto Nazionale di Fisica Nucleare, Laboratori Nazionali di Legnaro,*
 *Via Universitá 2, Legnaro (PD), Italy*

[d] *Physics Department, University of Catania,*
*Via S. Sofia 64, Catania, Italy*

*E-mail:* schillacif@lns.infn.it

ABSTRACT: A system of permanent magnet quadrupoles (PMQs) is going to be realized by INFN-LNS to be used as a collection system for the injection of laser driven ion beams up to 60 MeV/u in an energy selector based on four resistive dipoles. This system is the first element of the ELIMED (ELI-Beamlines MEDical and Multidisciplinary applications) beam transport, dosimetry and irradiation line that will be developed by INFN-LNS (It) and installed at the ELI-Beamlines facility in Prague (Cz). ELIMED will be the first user's open transport beam-line where a controlled laser-driven ion beam will be used for multidisciplinary researches. The definition of well specified characteristics, both in terms of performances and field quality, of the magnetic lenses is crucial for the system realization, for the accurate study of the beam dynamics and for the proper matching with the magnetic selection system which will be designed in the next months. Here, we report the design of the collection system and the adopted solutions in order to realize a robust system form the magnetic point of view. Moreover, the first preliminary transport simulations are also described.

KEYWORDS: Acceleration cavities and magnets superconducting (high-temperature superconductor; radiation hardened magnets; normal-conducting; permanent magnet devices; wigglers and undulators); Accelerator Applications; Beam Optics; Beam dynamics

---

[1] Corresponding author.

## Contents

## 1 Introduction

Laser-driven ion beams are a promising alternative to conventionally accelerated particle beams [1–5] even if they are not directly suitable for most applications because of the large angular and energy spread. Several efforts have been already done in order to develop beam-transport line able to produce a controllable beam from laser accelerated particles [6–11]. In 2014 the the FZU (ELI-Beamlines) launched a public tender to realize the beam transport, the dosimetric and the irradiation sections of the ELIMAIA (ELI Multidisciplinary Application of laser-Ion Acceleration) beam-line and the INFN-LNS has been officially appointed through a three years contract for its delivery. ELIMED will represent the section of ELIMAIA addressed to the transport, handling and dosimetry of laser-driven ion beams and to the achievement of stable, controlled and reproducible beams that, in the future, will be available for users interested in multidisciplinary and medical applications of such innovative technology. The transport beam-line that will be installed at ELIMAIA will be made of three elements: a collection system, namely a set of permanent magnet quadrupoles (PMQs) that will be placed close to the laser-interaction point, an energy selection system (ESS) based on four resistive dipoles, and a set of conventional electromagnetic transport elements. The beam-line will be working for laser-produced beams up to 60 MeV/u, offering, as output, a controllable beam in terms of energy spread (varying from 5% up to 20% for the highest energies), angular divergence and hence, manageable beam spot size in the range 0.1–10 mm and acceptable transmission efficiency (namely $10^6$–$10^{11}$ ions/pulse). In order to fulfill the project requirements, the two main elements of the beam-line, the collection system (PMQs) and the ESS, have to be optimized. The aim of the collection system is to collect the accelerated ions within a certain energy range, correct the angular divergence of the beam and efficiently inject it in the selection system. The beams coming out from this first part of the beam-line (PMQs+ESS) will have characteristics closer to conventional beams and, hence, easier to be transported and shaped with conventional magnetic lenses, such as resistive quadrupoles and steerers, which will be placed in the last part of the in-vacuum beam-line. The above description of the beam-line, makes it clear that the ESS is the core element around whom all the other magnetic devices have to be designed and realized, in particular the collection system has to be designed in order to properly inject the particles within a certain energy range in the magnetic chicane in order to ensure a good selection and transimission efficiency. Magnetic lenses are used in conventional particle accelerators to deflect, focus and/or correct the beam along the transport lines. PMQs lenses have the advantage to be relatively compact with an extremely high field gradient, of the order of 100 T/m, within a reasonable big bore of few centimeters. A PMQs system allows to collect most of the particles with wide divergence produced in the laser-target interaction process, providing a beam of good quality in terms of controlled size and divergence. For these reasons the interest in the application of PMQs for the beam handling of laser produced beams is growing in recent years [13–15]. Several PMQs designs have been proposed and developed, they are based on pure Halbach scheme [16] or hybrid devices using saturated iron to guide the magnetic field [17, 18]. Moreover, PMQs can be placed in the vacuum chamber, which means close to the laser-target interaction point, allowing a good collection and transmission efficiency even if the beam angular spread is of few degrees. In the following, a brief sketch of the selection system will be proposed in order to retrieve the conditions on the PMQs system transfer matrix necessary to match the two systems. These conditions have to be satisfied for a wide energy range of particles and are used to define the number of PMQs as well as the required field gradient. The free parameters to tune the system for each energy component are the relative distances among quadrupoles which depends on the maximum available space in the interaction chamber (1.2 m) and on the interaction forces between permanent magnet which fix the minimum space allowed between two PMQs. The design of the PMQs collection system is described in details in this paper together with the solutions adopted to solve demagnetization issues and magnets interaction forces. Preliminary optics simulation are also reported, showing the good performances of the system.

## 2 The Energy Selection System; preliminary conceptual design

A sketch of the ESS is shown in figure 1. It consists of four resistive dipoles with alternating fields, similar to a bunch compressor scheme; its main trajectory parameters are also reported in the figure, according to the description proposed in [12].

Using the main trajectory parameters shown in figure 1, it is possible to analytically define the characteristics of the ESS, in terms of radial and longitudinal dimensions of each dipole, field strength and selection slit aperture size, necessary to have the desired energy resolution. In fact, the bending radius $r$ and the bending angle $\alpha$ of a dipole are both function of the particle momentum. Hence, the energy can be written using the following relativistic equation:

$$E = c \sqrt{\frac{qB(\Delta x - d \tan \alpha)}{2(1 - \cos \alpha)}}^2 + M_0^2 c^2 - E_0 \qquad (2.1)$$

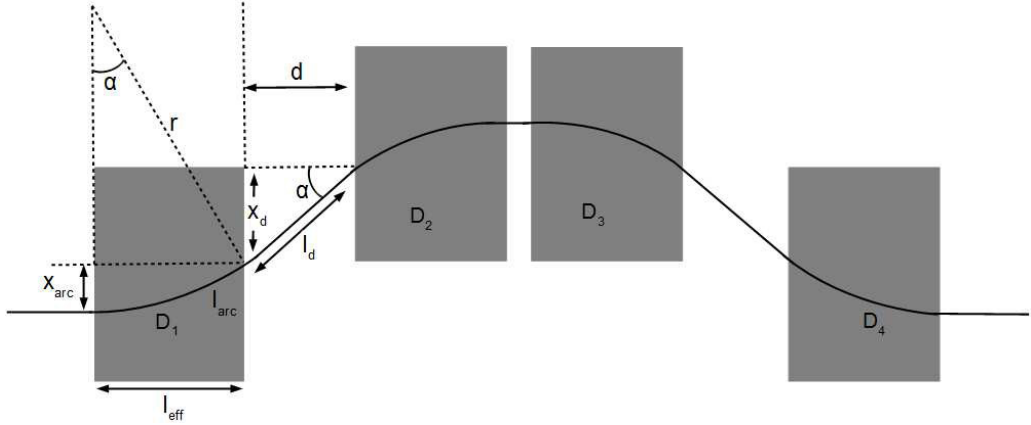

**Figure 1**. Energy Selection System layout and main trajectory parameters.

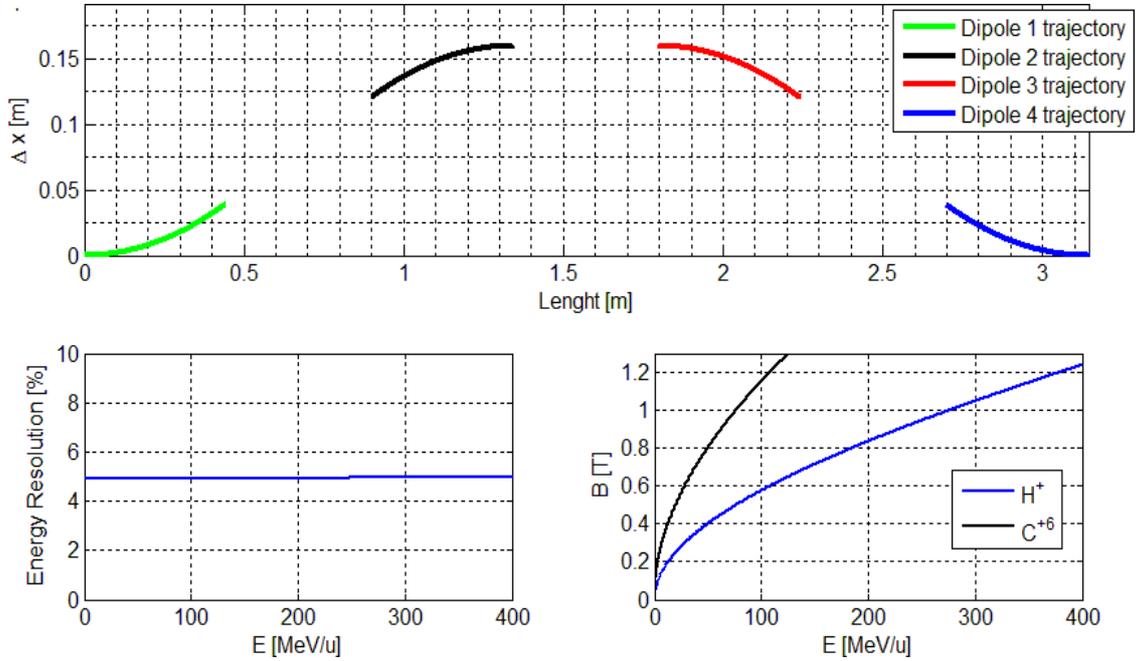

**Figure 2**. ESS preliminary characterization. The upper panel shows the design trajectory, the color arcs represent the path in the dipoles. The bottom panel shows the energy resolution for a fixed slit aperture size of 5 mm (r.h.s.) and the magnetic field as a function of the energy for protons and carbon ions (l.h.s.).

which represents the calibration equation of the ESS, being $\Delta x$ the radial displacement of the beam at the center of the device, $d$ the drift space between dipoles, $B$ the magnetic field, and $q$, $M_0$ and $E_0$ the characteristics of the ion to be selected, namely charge, rest mass and rest energy. The derivative of the previous equation times the slit aperture size $s$, namely $s*(dE/dx)$, gives the energy resolution of the device. Different layouts have been investigated and the best solution, in terms of compactness and performances, is the use of a single trajectory based device, as shown in figure 2.

For this system each magnet is 400 mm long (effective length of 450 mm) and the drift between each magnet is 500 mm long, for a total length of 3.1 m. The reference trajectory has maximum radial deflection of 160 mm at the device center where the selection slit is set. The upper panel of figure 1 shows the trajectory within the four dipoles (colored lines). The reference trajectory guarantees a fixed energy resolution of about (5 ± 0.05)% if a 5 mm aperture slit is used. The resolution does not depend on the particle energy and ion species, as shown in figure 2 left hand side of the bottom panel. In order to put particles with different energies on the reference trajectory the magnetic field of the system the magnetic field has to be changed, according the plot in figure 2 right hand side of the bottom panel. The plot shows that the field has to vary between 0.1 up to 1.2 T to select protons with energy ranging between 3 and 300 MeV and carbon ions up to 100 MeV/u. The proposed layout would allow to vary the energy resolution changing the slit aperture size, which is an advantage as at higher energies, laser-produced particles are less abundant and a bigger slit is necessary to keep the transmission efficiency acceptable. Moreover, the slit aperture size has to vary up to 20 mm to have an energy spread of the 20%. A system based on a single reference trajectory results more compact as the good field region can be reduced to 100 mm and dipoles can be C-shaped allowing an easy access to the vacuum chamber. Moreover, C-shaped dipoles can be displaced out of the main beam axis allowing the reduction of the magnets dimensions and the required magnetic field strength, which means an easier design of the dipoles especially in terms of iron saturation. The system is completed with two collimators with a diameter of 30 mm. The first collimator, set 200 mm upstream the device, is necessary to avoid spatial mixing of particles in the ESS that can considerably reduce the chicane energy resolution. The second collimator is set 200 mm downstream the ESS and refines the selection at the exit of the system. The detailed design of the ESS will be reported elsewhere.

## 3 Permanent magnet quadrupoles design

The described ESS layout has been used to calculate the main beam parameters to be considered in order to properly and efficiently inject the beam into the chicane, which basically depend on the effective trajectory length, 3.168 m, on the collimators size and the slit aperture size which is fixed to 40 mm on the transverse plane and variable on the radial (or dispersion) plane as described above. The knowledge of the reference path length allows to simplify the scheme of the ESS for the PMQs features definition. In fact, the chicane can be described as a three collimators system: the first one is 2,05 m downstream the target, the selection slit is far 1.584 m from the first aperture, namely at the middle point of the effective path length, the third collimator is set at the end of the path length. The optimal injection is obtained if the beam energy component to be selected has no divergence on the transverse plane and has a waist in correspondence of the selection slit in the radial plane. Hence, considering that the laser-produced beam has dimensions of a few tens of micron at the source and the slit aperture size is set to 1 mm to produce the narrowest energy spread of 1%, the collection system has to be a point-to-point focusing system on the radial plane and a point-to-parallel focusing system on the transverse plane. These requirements are satisfied if the transfer matrix elements of the system respect the conditions $M_{1,2} = 0$ and $M_{4,4} = 0$ [19] for a reference monochromatic beam ($H^+$ and $C^{6+}$) with energy ranging from 3 up to 60 MeV/u and 10° divergence (half angle) at the production point. A series of simulations has been performed to define the geometric and magnetic features of the collection system and, as a results, five quadrupoles, as described in the next table 1, are required. In this way the transfer matrix conditions can be satisfied for different energies changing the number of lenses used and their relative distances.

Table 1. PMQs main features.

| n° of PMQs | Geometric Length | Field Gradient | Bore Diameter |
|---|---|---|---|
| 1 | 160 mm | 101 T/m | 30 mm |
| 2 | 120 mm | 99 T/m | 30 mm |
| 2 | 80 mm | 94 T/m | 30 mm |

Once the PMQs features have been defined, the matching between the PMQs and the linearized ESS have been optimized changing the quadrupoles relative distances, which are the free parameters of the system, and using as constraints the Twiss Parameters and the emittance of a beam passing with no losses through the three collimators of the linearized chicane. This beam parameters depends on the reference beam energy to be transported and also on the aperture slit size used, namely 5 mm for the lower energy up to 20 mm for higher energy beam component. In the next chapter the layout of the collection system for the injection of 60 MeV protons in the ESS is reported.

The PMQs system has to be versatile as it has to be tuned for the injection in the ESS of reference energies varying in a wide range and to ensure a reasonably good transmission efficiency. Thus, a big bore of at least 36 mm is necessary with a strong field gradient and high uniformity within, the 75% of its surface. The net bore has to be reduced to 30 mm in diameter as a 3 mm thick shielding pipe for magnet protection has to be set in the aperture. Considering these requirements the quadrupoles design is based on a standard Halbach array with trapezoidal magnets surrounded by two external hybrid arrays made of rectangular permanent magnet blocks and iron. The inner array of 122 mm outer diameter is mainly responsible for the field quality of the quadrupole, the external arrays, of 223 mm and 322 mm respectively, are necessary to increase the volume of permanent magnet material and to reach the required field gradient. A scheme of the PMQs layout is shown in figure 3.

The choice of this layout, in which a pure Halbach array is mixed with hybrid Halbach arrays, is due to the fact that permanent magnet alloys are very brittle material and the realization of trapezoidal blocks is problematic from the mechanical point of view. This design, as shown in this section, results to be robust with a very good field quality and, at the same time, easier to realize, compared to a pure Halbach array. As can be seen in figure 3 the two external arrays are set within an iron frame which has the function of supporting structure as well as magnetic flux guide. The inner array is made of two different permanent magnet alloys with different characteristics. This is due to the fact that the magnetic flux in the bore has a maximum value of about 1.8 T and, for such high flux values, local demagnetization phenomena can occur, where the magnetic field components Hx and Hy have opposite direction respect to the permanent magnets magnetization direction and a value higher than the material coercitivity. This issue can be solved mixing materials with different coercitivity in some strategic places of the array. In particular in the proposed design the main magnetic material is *NdFeBN* 48*H* with remanence $B_r$ = 1.39 T and coercitivity $Hc$ = 1273 kA/m and *NdFeBN* 38*UH* ($Br$ = 1.26 T and $Hc$ = 1990 kA/m), the high coercitivity component, set in the point at risk of demagnetization. Figure 4 shows a particular of the inner array with the sectors where the $H$ field components are higher than the main magnetic material coercitivity, the red arrows are the $H$ field directions and the blue arrows are the magnetization direction of each block.

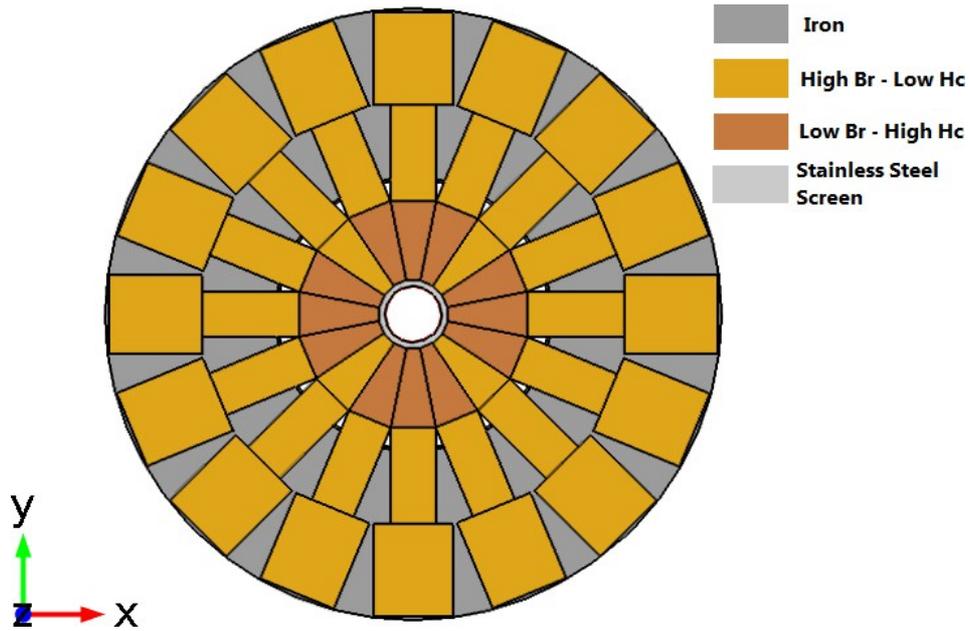

**Figure 3**. PMQ layout. The different colors are referred to the different materials used.

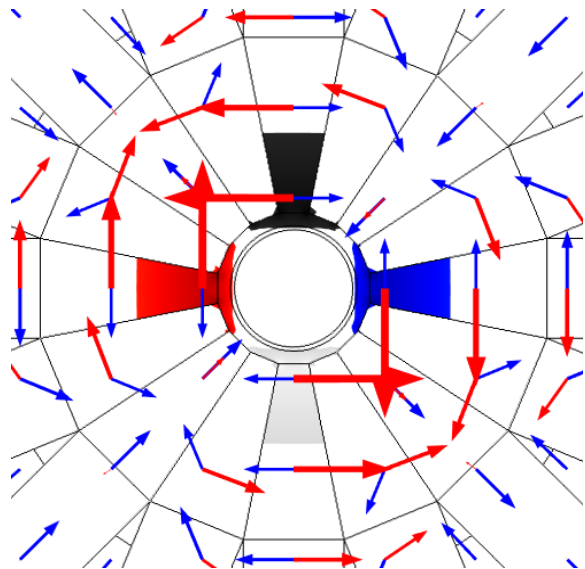

**Figure 4**. Particular of the inner array with demagnetization field components: colored zones have a H field higher than NdFeB N48H coercitivity, red arrows are in the H field directions and blue arrows are in the magnetization directions of each block.

Comparing figure 3 and figure 4, it is clear that the parts of the magnetic array at risk of demagnetization are made of high coercitivity material. The volume occupied by the high coercitivity material is, anyway, bigger than necessary as a safety margin of 30% has been considered.

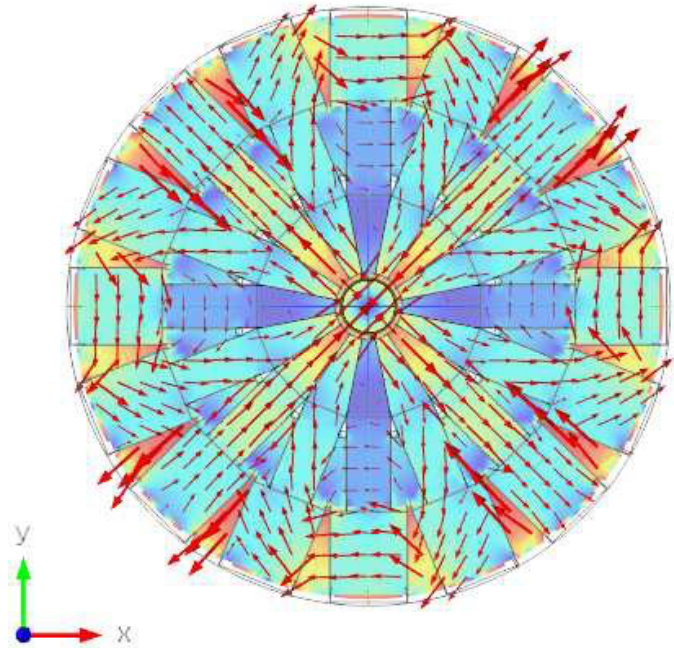

**Figure 5**. Magnetic flux density (colour surfaces) and magnetization directions (red arrows).

The maximum value of the $H$ field is 1.8 kA/m, this value determines the choice of the secondary magnetic alloy. The permanent magnet materials have been simulated by means of their BH curves with a stable behavior in the temperature range between 20 °C and 60 °C.

### 3.1 PMQ field analysis

The field analysis is here limited to the long (160 mm) quadrupole and within the active bore of 30 mm diameter. As described above, each PMQ can be seen as the combination of a 16 sectors pure Halbach array and two 16 sectors hybrid arrays. Each sector has its own magnetization direction as shown in figure 5. The maximum value of $B$ in the net bore (30 mm diameter) is 1.55 T, the Halbach formula [20] for a similar ideal quadrupole, gives a $B_{max}$ = 2 T, and, considering the differences between the present design and the ideal one, the agreement can be considered satisfactory.

The maximum gradient is of about 101 T/m and its uniformity is within 2% at 12 mm from the quadrupole center over 360°, see figure 6. This distance represents the good field region radius. The integrated gradient uniformity is reported in figure 7. The field uniformity has been evaluated as described in [18] and resumed in table 2 where a comparison with an ideal Halbach array is also presented.

The effective length of the quadrupole is 163 mm, for both the presented design and the pure ideal Halbach design, and the focal distance ranges from 15.2 mm, for 3 MeV protons, up to 69.1 mm, for 60 MeV protons. The harmonic content is also important and have to be known and controlled in order to minimize non-linear effects on the beam dynamics. Beside the second field harmonic, which is the quadrupole component, the proposed hybrid design has a very small harmonic content and it is again similar to a pure Halbach design. The normal harmonic content is basically the same for both pure and hybrid (0.31% of the main harmonic), the complex harmonic content is slightly different: 0.38% for a pure Halbach quadrupole and 0.39% for the hybrid design. These results show that the proposed design has basically the same features of a pure Halbach quadrupole resulting, on the other hand, a cost effective solution.

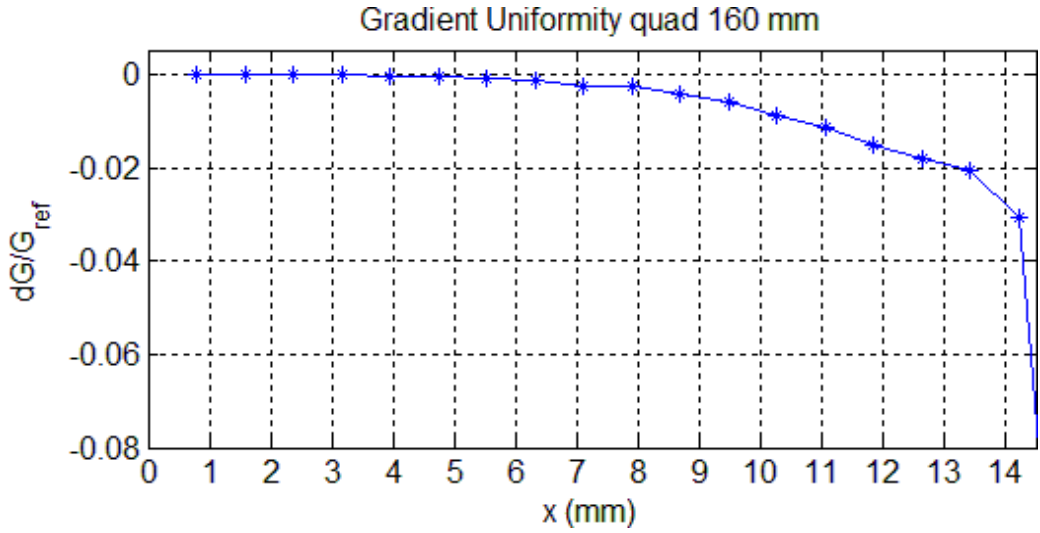

**Figure 6**. Gradient Uniformity for the 160 mm PMQ.

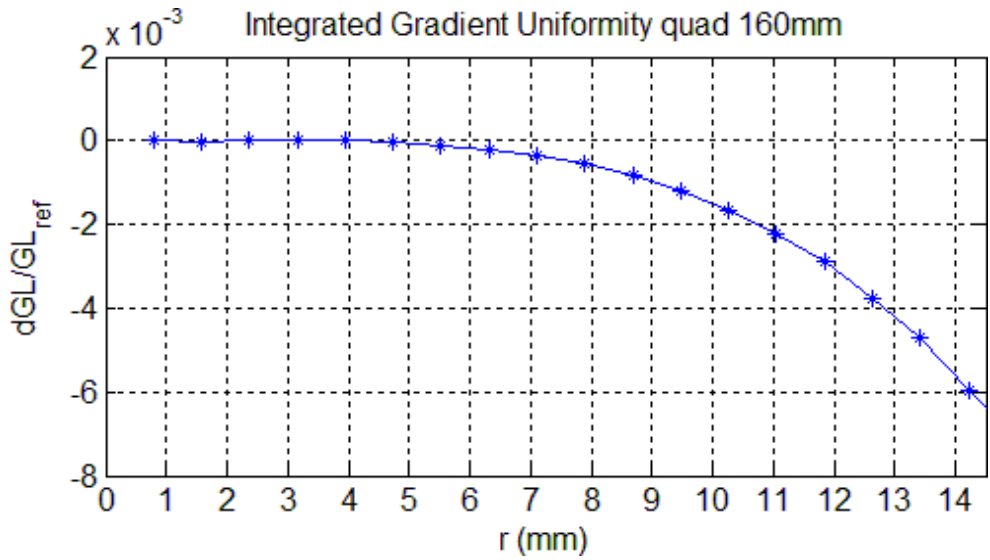

**Figure 7**. Integrated Gradient Uniformity for the 160 mm PMQ.

Table 2. Field Uniformity of the 106 mm PMQ compared with a corresponding pure Hablach design.

|  | Gradient Uniformity | | Integrated Gradient Uniformity | |
|---|---|---|---|---|
|  | PMQ 160 mm | Pure Halbach | PMQ 160 mm | Pure Halbach |
| @ $R$ = 7.5 mm | < 0.25% | < 0.1% | < 0.035% | < 0.03% |
| @ $R$ = 12 mm | < 2% | < 1.5% | < 0.3% | < 0.3% |
| @ $R$ = 14.5 mm | < 8% | < 3% | < 0.7% | < 0.6% |

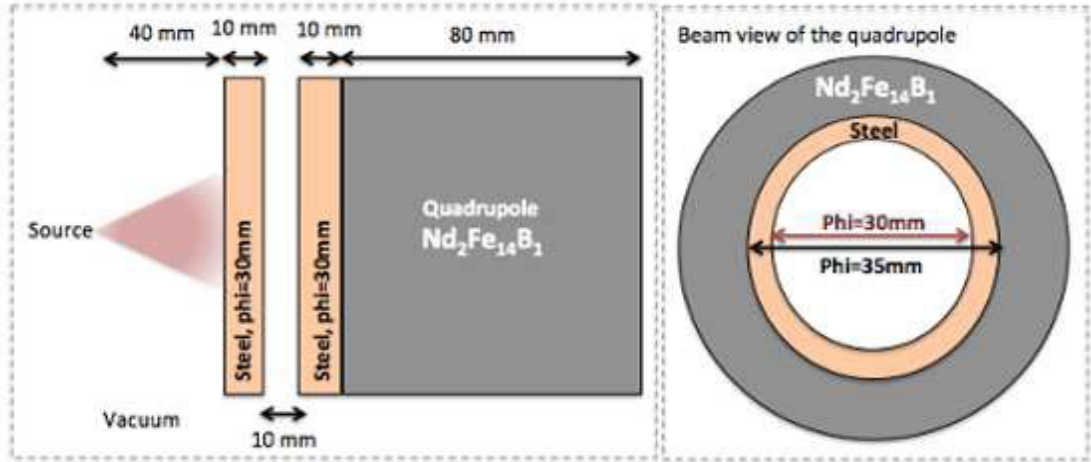

**Figure 8**. Set-up of GEANT4 the simulations.

### 3.2 Thermal stability and secondary neutron flux

Thermal stability and secondary neutron flux are usually issues for NdFeB permanent magnet alloys and the GEANT4 toolkit has been used for simulating the behavior of the magnetic material in realistic laser-target interaction environment. The geometrical set-up of the simulations is shown in figure 8. A stainless steel shield, set in front of the first quadrupole and necessary for the protection of the whole system (especially mechanicals components) is placed downstream the source. After a small drift, the quadrupole is placed considering its own protection shield, a 10 mm thick stainless steel plate, and the inner shielding pipe, which have been considered 2.5 mm thick for these simulations. The PMQ material has been simulated using the NdFeB density ($d$ = 7.4 g/cm$^3$) and its specific heat (440 J/(kg $\star$K)).

The spectrum at the source is a typical TNSA exponential distribution, normalized in order to have $10^9$ particles at 60 MeV $\pm$ 5% and 10° of half angle divergence. The maximum temperature increase results to be $10^{-3}$ K on the external layer of the PMQ shielding pipe. Therefore, to be conservative, the inner quadrupole surface, in contact with the pipe, has been considered as to have the same rise in temperature. In order to increase the total temperature from 20°, room temperature, to 60° ($\Delta T$ = 40°C which is the stability range of the material used in the PMQs), the total number of possible shot is of about 20000. This means that the PMQs can be safely used for 0.5 hour if the laser repetition rate is 10 Hz and for 5 hours in the 1 Hz repetition rate case. This operational time is still conservative as the relaxation time between each particle bunch has not been taken into account [21].

The neutron flux have been evaluated using the Bertini Cascade model implemented in GEANT4 (QGSP_BERT) [22] which can simulate the hadronic processes in our energy range with a very good agreement respect to the experimental data. The maximum neutron production is $10^2$ neutrons/cm$^2$. In literature, radiation damage of permanent magnet is reported to occur for higher neutron fluxes [23, 24]. The simulation outputs, are shown in figure 9. The upper panel resumes the results of the temperature increasing on three different transverse planes of the quadrupole. The neutron flux is shown in the bottom panel. Anyway this preliminary results will be reviewed with more realistic laser-driven bunches, in terms of angular divergence.

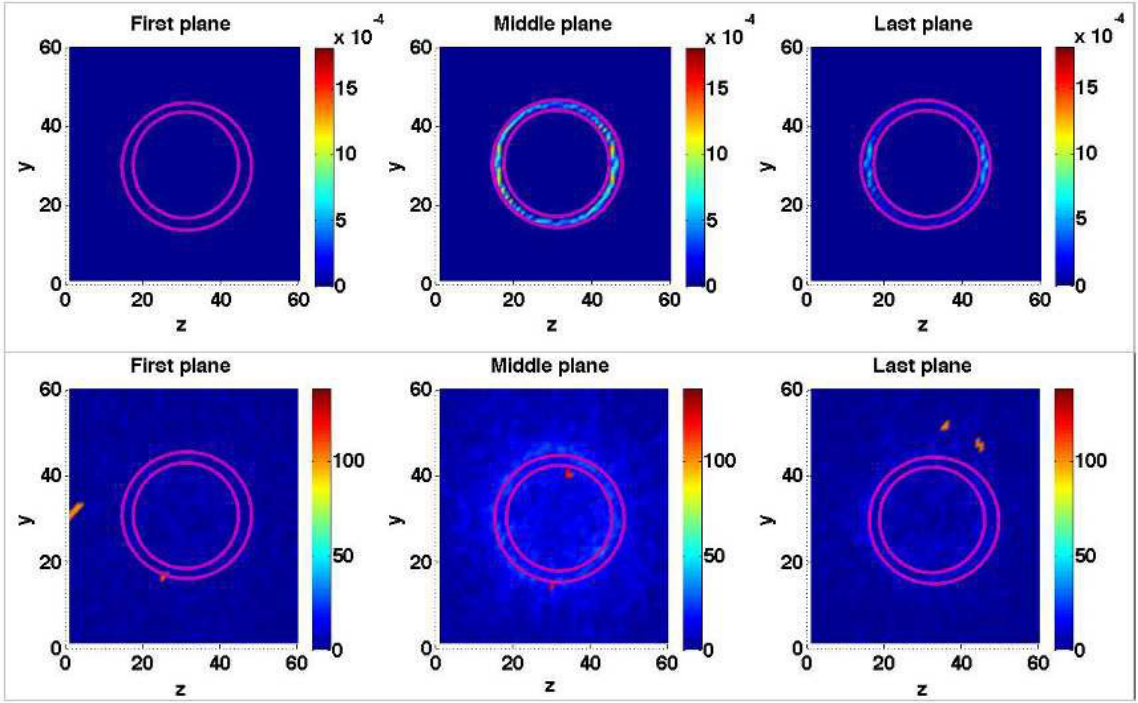

**Figure 9**. Temperature increasing in Kelvin (upper panel) and neutron production in *neutrons/cm$^2$* (bottom panel) for the first, middle and last plane of the quadrupole. Magenta circles delimit the inner shielding pipe.

### 3.3 Attraction/repulsion forces between PMQs

One more issue related to permanent magnets is due to the attraction/repulsion forces that can be extremely high when large volumes are involved. Hence, considering that the PMQs system won't be static as it as to be tuned for a certain energy, changing the relative distances between magnets and, also, the number and polarity of each quadrupole can be changed, it is important to evaluate these forces for the proper design of the mechanics and also to define a safe procedure for the PMQs extraction and/or assembly. The forces have been evaluated between the 160 mm length PMQ and one of the 120 mm PMQs, which represents the case with biggest volume of permanent magnets are interacting. The results have been benchmarked performing a comparison of the same calculations on a set of small PMQs already realized [18] and the measured forces. The calculations have been performed changing the relative distance between the quadrupoles from 22 mm (the minimum distance possible due to the 10 mm think stainless steel screens that will be set at the front and at the bottom of the quadrupoles) up to 192 mm. Results show that the force are extremely high for small distances (about 16000 N) but they decrease quickly and are reasonably small at distances larger than 100 mm. This is important to identify the minimum working distances of the PMQs, in order to realize the proper mechanics, which has been fixed at 40 mm, corresponding to a force of about 5 kN, and the minimum distance for the extraction of the magnets, which has been fixed at 180 mm corresponding to a force smaller than 80 N.

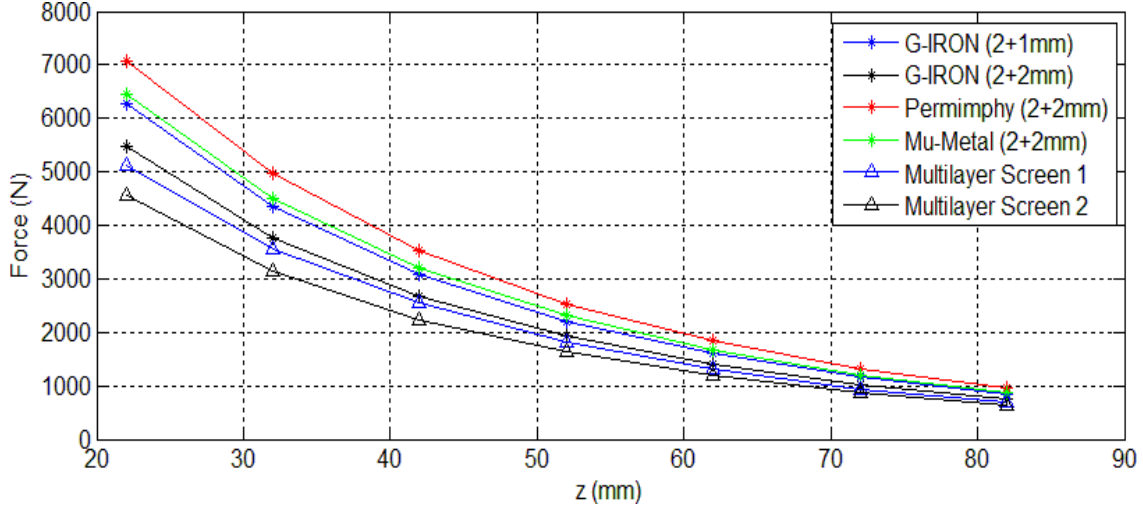

**Figure 10**. Force between two quadrupoles at different distances and with different screens.

The transport efficiency of the system depends on the minimum distance between each PMQ, as the mentioned condition are well satisfied if the geometric constraints are less strict. Hence, different possible shields for the magnetic field responsible of these high forces have been investigated in order to reduce it. In general non-linear materials such as Mu-metal are used to shield the magnetic fields but they usually have low saturation and high permeability and are not efficient for this case. Different alloys such as G-IRON or MAGSHIELD [25, 26] offering high saturation and relatively low permeability, result to be more suitable for this application. A comparison of the shielding efficiency among different alloys and multilayer screens is shown in figure 10. G-IRON and MAGSHIELD give very similar results in the simulations as they have almost the same magnetic features, and, for this reason, only the results with G-IRON are here reported.

The multilayer screens are made of G-IRON foils alternating with iron foils of 2 mm thickness for case 1 and 1 mm thickness for case 2. It should be noted, from figure 10, that for distances higher than 75 mm all the different materials investigated have almost the same shielding efficiency. This is due to the fact that the total field intensity is smaller and the non-linear materials are not saturated. This also justifies the necessity to use materials with high magnetic saturation. The results obtained with G-IRON shields have been benchmarked on the prototypes already developed confirming the simulations results and the possibility to reduce the minimum distance to 30 mm improving the transmission efficiency of the PMQs.

## 4 Beam-transport simulations

In this section the preliminary beam-transport simulations results are reported. The purpose of these simulations is to optimize the PMQs to inject a monochromatic beam component in the

– 11 –

Table 3. PMQs set up for the collection of 60 MeV protons.

| PMQ length | Gradient | Distance from target |
|---|---|---|
| 160 mm | −101 T/m | 54 mm |
| 120 mm | 99 T/m | 258 mm |
| 120 mm | −99 T/m | 423 mm |
| 80 mm | 94 T/m | 748 mm |
| 80 mm | −94 T/m | 872 mm |

energy selector sketched as a three collimators system, geometrically placed along the beam-line as described in section 2.

Figure 11 shows the scheme of the beam-line. The laser-target interaction point is on the axis origin, the PMQs system and then the energy selection system follow. D1 and D3 are the collimators, D2 is the selection slit. The simulated beam has been considered monochromatic in a first phase, with 40 $\mu$m diameter circular shape at the production point and an uniform angular divergence of 11° (half angle). Simulations have been performed in the energy range between 3 and 60 MeV for protons and for $C^{+6}$ up to 60 MeV/u. Here, only the cases with maximum energy are described. In the simulations, the field maps of the quadrupoles described in the previous section are used. The goal of this preliminary simulations is the optimization of the PMQs, which has to stay confined in 1200 mm from the source, in order to inject the beam in the energy selector avoiding the losses downstream the collection system and in the linearized chicane. The optimal injection can be obtained having a beam parallel (or with a small divergence of few *mrad*F) on the transverse plane (bottom panel of figure 11) and a waist close to the selection point on the radial plane (upper panel of figure 11). The PMQs system have been optimized changing the relative distances between magnets in order to respect the transfer matrix and beam parameters requirement described in the previous chapter. Moreover, from the geometric point of view, other constraints have been considered: the minimum allowed distance from the laser-target interaction point is 50 mm and the minimum allowed distance between each quadrupole is 40 mm, according the calculated forces without shield reported in the previous section. The optimized system is described in table 3.

### 4.1  60 MeV protons case

The beam envelope is shown in figure 11. The monochromatic simulations gives a transmission efficiency of 10%, with the losses limited in the PMQs system. If a guassian angular distribution is considered, with a FWHM of 5° half angle, the transmission efficiency raise up to the 15%. The output beam is shown in the phase space plot in figure 12. As it can be seen the beam divergence is considerably reduced to 0.3°, which makes it possible to transport and shape the beam using conventional quadrupoles. The filamentations of the beam are due to the natural field uniformity of the PMQs.

For this energy case, simulations with more realistic TNSA protons have been considered. The input spectrum has the typical exponential shape with a cut-off at 110 MeV and the beam divergence is uniformly distributed between 0° and the maximum angles associated to each energy, as shown in figure 13, where the red dots represents the function used to fit the data and reproduce the angles. The plot represents the half-angle distribution.

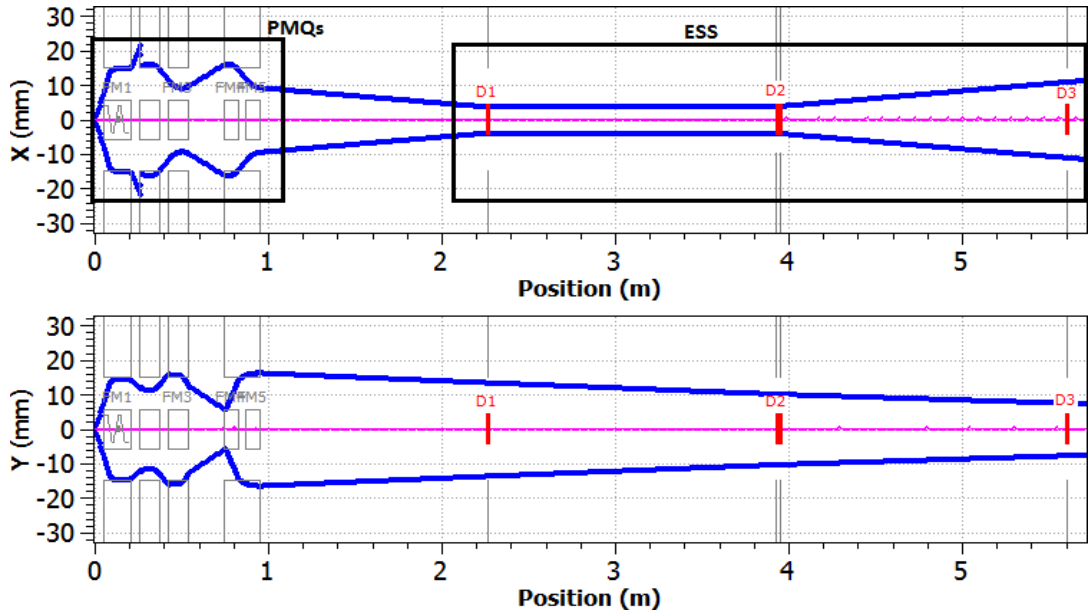

**Figure 11**. Scheme of the simulated beamline with envelopes in the radial (upper panel) and transverse (bottom panel) planes of 60 MeV protons.

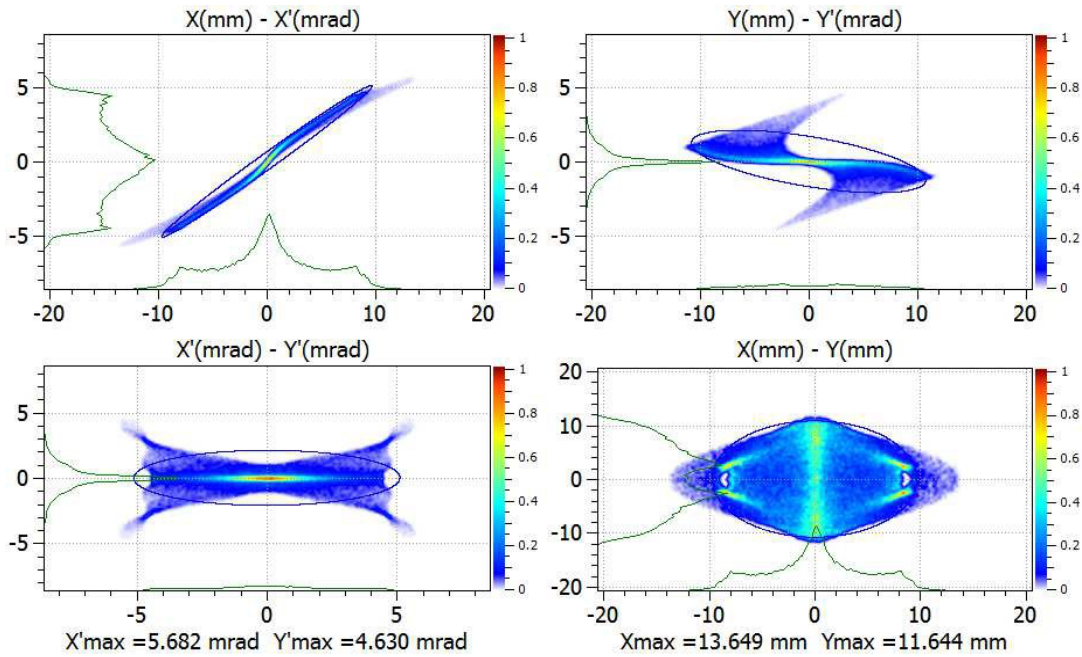

**Figure 12**. Phase space plot of the 60 MeV monochromatic proton beam downstream the ESS.

In figure 14 the red line represents the laser-accelerated spectrum, the black line is the spectrum transmitted by the PMQs system, the blue line is the spectrum at the selection plane and the green line is taken at the ESS exit, simulated as a collimators system. What is important from this simulations is the fact that for the energy under consideration there is no losses in the energy range

– 13 –

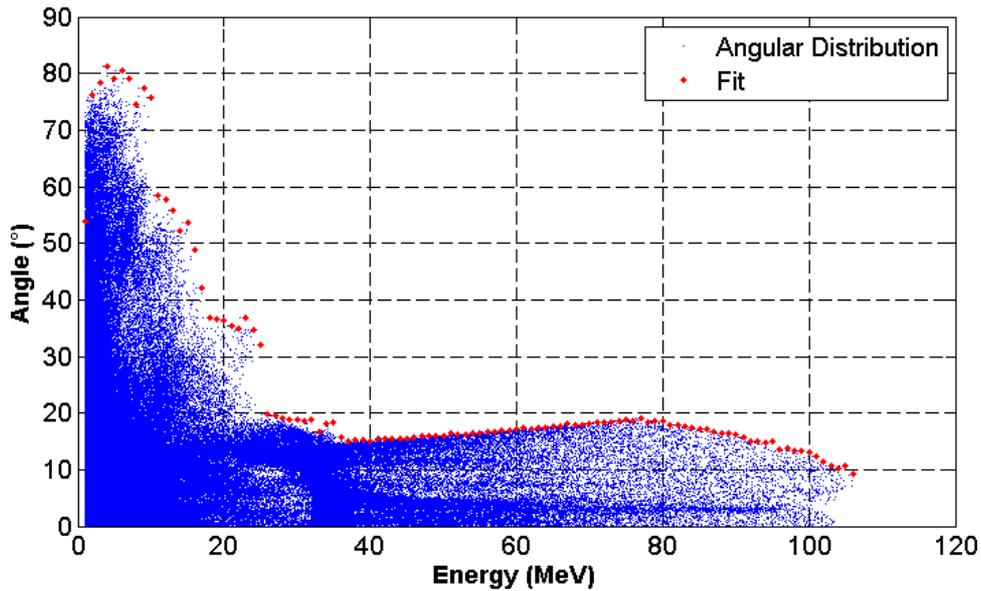

**Figure 13**. Angular distribution (blue dots) and function used in the simulations (red dots).

of interest (60 MeV with a spread of 10%) apart from those in the PMQs system and the transmission efficiency in this range is of about 4%. This ensures a minimum number of particles of about $10^7$ per pulse in the worse case. Moreover, the spectrum is not cleaned up by the ESS magnetic field, which justify the broad energy spread at the beam-line output. It should be noted that the collection efficiency seems to have a maximum for 30 MeV protons; this is due to their higher abundance with respect to the 60 MeV component and to their magnetic rigidity. In fact, the optics in the quadrupole is similar for 30 and 60 MeV, but the matching between PMQs and ESS is not perfect for 30 MeV with the presented configuration and, as can be seen from the plot, this component has losses exceeding the 30% in the chicane. Anyway, such an output spectrum can be selected and the possibility to realize the ESS magnets with laminated iron, instead of using a solid core, would permit the possibility of fast changes in the magnetic field and to select beams with different central energies with a reasonably high amount of particles, even if the PMQs is not set up for those energies and the transmission efficiency is not at its maximum.

### 4.2  60 MeV/u carbons case

The system performances have been studied for other ion species and results for monochromatic $C^{+6}$ with energy of 60 MeV/u are here reported. The beam envelope is shown in figure 15. Four quadrupoles are required to inject the beam in the ESS. The angular spread is limited to $0.25°$.

## 5  Conclusions

The PMQs system have been designed to have high field quality and all the possible issues related to the magnetic design have been investigated and solved, such as demagnetization issue, the thermal stability and the secondary neutron flux. The attraction/repulsion forces between quadrupoles are crucial for the transport efficiency and the shielding effectiveness of the non-linear materials considered in the simulations will be tested in order to review the optics of the system with more relaxed constraints on the relative PMQs distances. The simulations results on the beam transport have to be considered preliminary as the divergence of the beam is worse than the one coming from the PIC simulations output. In fact, each energy component has an uniform divergence between ±θ, being θ the maximum angle of figure 13. More realistic simulations would have to use a gaussian angular distribution. Moreover the work on the Energy Selection System is still ongoing and the matching between the two systems is not very well optimized at this stage. Due to this approximations, it is reasonable to suppose that the transmission efficiency of the beam-line can be increased.

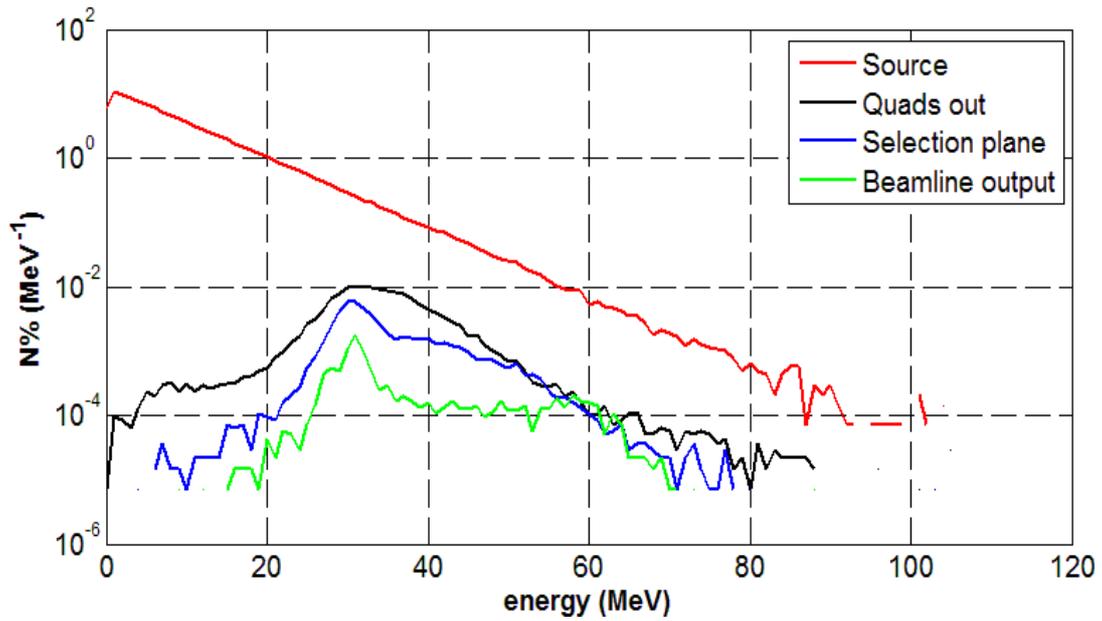

**Figure 14**. Particle spectrum at different positions along the beamline.

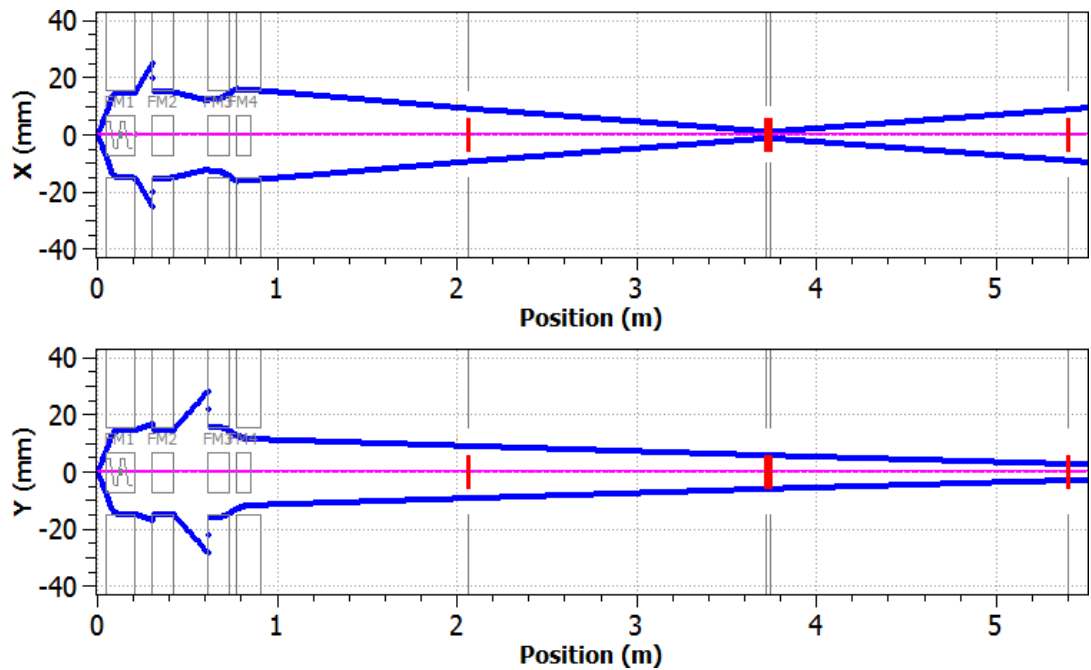

**Figure 15**. Scheme of the simulated beamline with envelopes in the radial (upper panel) and transverse (bottom panel) planes of 60 MeV/u $C^{+6}$.

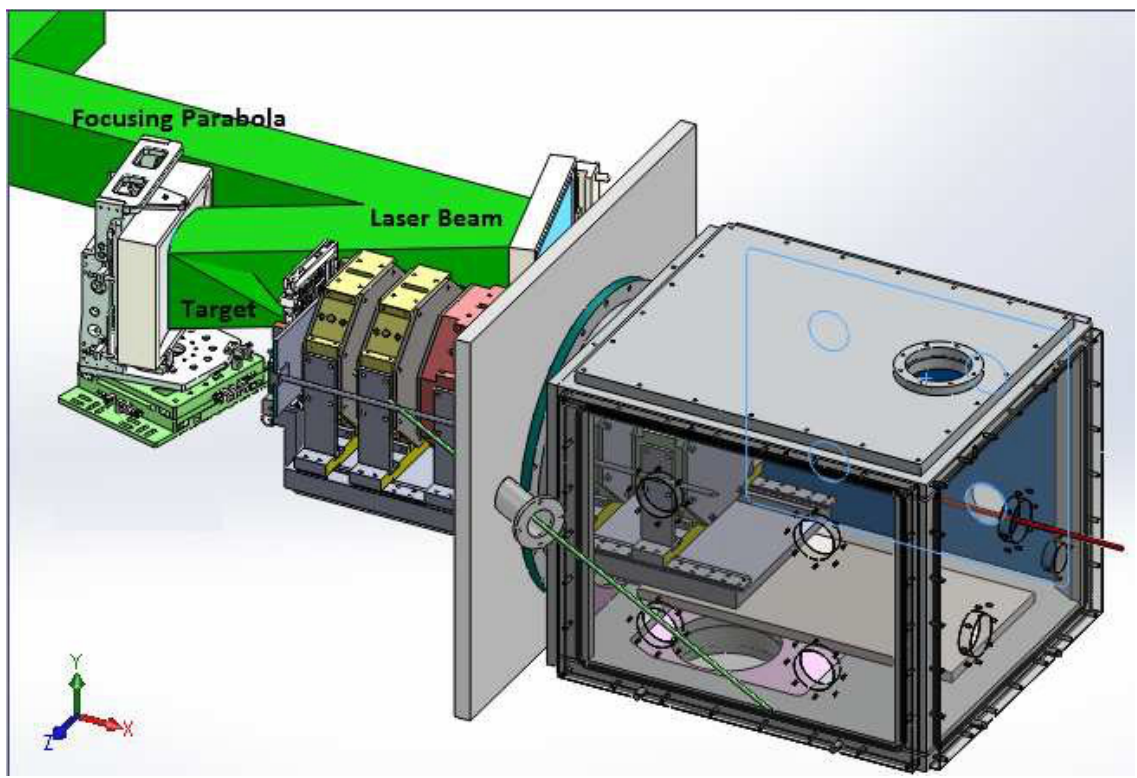

**Figure 16**. Scheme of the PMQs set on the beamline with related displacement system.

The PMQs system will be placed partially inside the main interaction chamber, left side of figure 16, in order to reduce the distance form the target, and partially inside and auxiliary chamber. The PMQs will be displaced on a linear guide where they can be displaced and they relative distances can be changed by means of proper vacuum motors. The mechanics also allow the extraction of the PMQs for the top, opening the main and the auxiliary chamber.

Moreover, as stated above, an electromagnet able to change field each second could allow to produce beamlets with a very narrow energy spread providing an active energy modulation system, which is crucial for the production of a spread out Bragg peak to demonstrate the clinical applicability of laser-driven ion beams.

# References

[1] V. Malka et al., *Practicability of protontherapy using compact laser systems*, *Med. Phys.* **31** (2004) 1587.

[2] S.V. Bulanov, T.Z. Esirkepov, V.S. Khoroshkov, A.V. Kuznetsov and F. Pegoraro, *Oncological hadrontherapy with laser ion accelerators*, *Phys. Lett.* **A 299** (2002) 240.

[3] T. Esirkepov, M. Borghesi, S.V. Bulanov, G. Mourou and T. Tajima, *Highly Efficient Relativistic-Ion Generation in the Laser-Piston Regime*, *Phys. Rev. Lett.* **92** (2004) 175003.

[4] A.V. Kuznetsov, T.Z. Esirkepov, F.F. Kamenets and S.V. Bulanov, *Efficiency of ion acceleration by a relativistically strong laser pulse in an underdense plasma*, *Plasma Phys. Rept.* **27** (2011) 211.

[5] F. Schillaci et al., *ELIMED, MEDical and multidisciplinary applications at ELI-Beamlines*, *J. Phys. Conf. Ser.* **508** (2013) 012010.

[6] V. Scuderi et al., *Development of an energy selector system for laser-driven proton beam applications*, *Nucl. Instrum. Meth.* **A 740** (2014) 87.

[7] M. Maggiore, G.A.P. Cirrone, F. Romano, A. Tramontana, F. Schillaci and V. Scuderi, *Transport and energy selection of laser produced ion beams for medical and multidisciplinary applications*, in proceedings of *The 5$^{th}$ International Particle Accelerator Conference (IPAC 2014)*, Dresden, Germany (2014), TUPME034 1425-1427 [ISBN: 978-3-95450-132-8], http://accelconf.web.cern.ch/AccelConf/IPAC2014/papers/tupme034.pdf.

[8] S. Busold et al., *Commissioning of a compact laser-based proton beam line for high intensity bunches around* 10 *MeV*, *Phys. Rev. ST Accel. Beams* **17** (2014) 031302.

[9] G.A.P. Cirrone et al., *ELIMED a new concept of hadrontherapy with laser-driven beams*, *IEEE Nucl. Sci. Symp. Med. Imag. Conf.* (2002) 1999.

[10] F. Schillaci et al., *ELIMED: medical application at eli-beamlines. Status of the collaboration and first results*, *Acta Polytechnica* **54** (2014) 285.

[11] G.A.P. Cirrone et al., *ELIMED, future hadrontherapy applications of laser-accelerated beams*, *Nucl. Instrum. Meth.* **A 730** (2013) 174.

[12] P. Castro, *Beam trajectory calculations in bunch compressors of FFT2*, TECHNICAL-NOTE-2003-01 (2003).

[13] S. Becker et al., *Characterization and tuning of ultrahigh gradient permanent magnet quadrupoles*, *Phys. Rev. ST Accel. Beams* **12** (2009) 102801.

[14] H. Sakaki et al., *Simulation of Laser-Accelerated Proton Focusing and Diagnosis with a Permanent Magnet Quadrupole Triplet*, *Plasma Fusion Res.* **5** (2010) 009.

[15] M. Schollmeier et al., *Controlled Transport and Focusing of Laser-Accelerated Protons with Miniature Magnetic Devices*, *Phys. Rev. Lett.* **101** (2008) 055004.

[16] K. Halbach, *Physical and Optical Properties of Rare Earth Cobalt Magnets*, *Nucl. Instrum. Meth.* **187** (1981) 109.

[17] T. Mihara, Y. Iwashita, M. Kumada, C.M. Spencer and E. Sugiyama, *Superstrong Adjustable Permanent Magnet for a Linear Collider Final Focus*, in proceedings of *The 12$^{th}$ Linear Accelerator Conference (LINAC04)*, Lubeck, Germany (2004), SLAC Report SLAC-PUB-10878, http://www.slac.stanford.edu/cgi-wrap/getdoc/slac-pub-10878.pdf.

[18] F. Schillaci, M. Maggiore, D. Rifuggiato, G.A.P. Cirrone, G. Cuttone and D. Giove, *Errors and optics study of a permanent magnet quadrupole system*, 2015 *JINST* **10** T05001.

[19] H. Wollnik, *Optics of Charged Particles*, Academic Press Inc., (1987).

[20] K. Halbach, *Design of Permanent Multipole Magnets with Oriented Rare Earth Cobalt Material*, *Nucl. Instrum. Meth.* **169** (1980) 1.

[21] P. Knaus and J.A. Uythoven, *Transient thermal analysis of intense proton beam loss on a kicker magnet conductor plate*, CERN-SL-2000-034-BT (2000).

[22] A. Bungau, R. Cywinski, R. Barlow, C. Bungau, P. King and J. Lord, *GEANT4 modeling of heat deposition into the isis muon target*, in proceedings of *2011 Particle Accelerator Conference*, New York, NY, U.S.A. (2011), TUP004, http://accelconf.web.cern.ch/accelconf/pac2011/papers/tup004.pdf.

[23] N. Simos, P.K. Job and N. Mokhov, *An experimental sudty of radiation-induced demagnetization of insertion device permanent magnets*, in proceedings of *The 11[th] European Particle Accelerator Conference (EPAC08)*, Genoa, Italy (2008), WEPC05, http://www.bnl.gov/isd/documents/43363.pdf.

[24] X.-M. Maréchal, T. Bizen, Y. Asano and H. Kitamura, *65 MeV neutron irradiation of NeFeB permanent magnets*, in proceedings of *The 10[th] European Particle Accelerator Conference (EPAC06)*, Edinburgh, Scotland (2006), THPCH13, https://accelconf.web.cern.ch/accelconf/e06/PAPERS/THPCH135.PDF.

[25] http://g-iron.it/.

[26] http://www.lessemf.com/mag-shld.html.